\documentclass[twocolumn]{aastex63}
\usepackage{amsmath}

\received{June 4, 2021}
\revised{September 27, 2021}
\accepted{September 30, 2021}
\submitjournal{AJ}

\usepackage{color}
\usepackage{xcolor}
\usepackage{mathtools}
\usepackage{CJKutf8}
\shorttitle{Dynamical Stability of Multi-Planet Systems}
\shortauthors{Wei et al.}

\graphicspath{{./}{figures/}}

\begin{document}
\begin{CJK*}{UTF8}{gbsn}

\title{Relativistic Dynamical Stability Criterion of Multi-Planet Systems with a Distant Companion}

\correspondingauthor{Smadar Naoz}
\email{snaoz@astro.ucla.edu}

\author[0000-0002-2612-2933]{Lingfeng Wei (魏凌枫)}
\affiliation{Center for Astrophysics and Space Sciences, University of California, San Diego, La Jolla, CA 92093, USA}

\author[0000-0002-9802-9279]{Smadar Naoz}
\affil{Department of Physics and Astronomy, University of California, Los Angeles, CA 90095, USA\\}
\affil{Mani L. Bhaumik Institute for Theoretical Physics, Department of Physics and Astronomy, UCLA, Los Angeles, CA 90095, USA\\}

\author[0000-0003-3799-3635]{Thea Faridani}
\affil{Department of Physics and Astronomy, University of California, Los Angeles, CA 90095, USA\\}
\affil{Mani L. Bhaumik Institute for Theoretical Physics, Department of Physics and Astronomy, UCLA, Los Angeles, CA 90095, USA\\}

\author[0000-0003-1540-8562]{Will M. Farr}
\affiliation{Department of Physics and Astronomy, Stony Brook University, Stony Brook, NY 11794, USA}
\affiliation{Center for Computational Astrophysics, Flatiron Institute, 162 Fifth Avenue, New York, NY 10010, USA}

\begin{abstract}

Multi-planetary systems are prevalent in our Galaxy. The long-term stability of such systems may be disrupted if a distant inclined companion excites the eccentricity and inclination of the inner planets via the eccentric Kozai-Lidov mechanism. However, the star-planet and the planet-planet interactions can help stabilize the system. In this work, we extend the previous stability criterion that only considered the companion-planet and planet-planet interactions by also accounting for short-range forces or effects, specifically, relativistic precession induced by the host star. A general analytical stability criterion is developed for planetary systems with $N$ inner planets and a relatively distant inclined perturber by comparing precession rates of relevant dynamical effects. Furthermore, we demonstrate as examples that in systems with $2$ and $3$ inner planets, the analytical criterion is consistent with numerical simulations using a combination of Gauss's averaging method and direct N-body integration. Finally, the criterion is applied to observed systems, constraining the orbital parameter space of a possible undiscovered companion. This new stability criterion extends the parameter space in which an inclined companion of multi-planet systems can inhabit.

\end{abstract}


\section{Introduction} 
\label{sec:intro}

Recent ground and space-based observations suggest that multi-planet systems are abundant in our galaxy \citep[e.g.,][]{Lissauer_2011, Lissauer_2012, Tremaine_2012, Rowe_2014}. Many of these systems contain multiple planets on tight orbits \citep[e.g.,][]{Howard_2012, Brewer_2018, Wu_2019}. The NASA Kepler Mission detected hundreds of multi-planet systems with semi-major axis smaller than $1$~au \citep[e.g.,][]{Borucki_2011}. Some systems even have multiple planets all residing within mercury's semi-major axis, such as the system of TRAPPIST-1 \citep[see][]{Gillon_2017}. Ultra-short period (USP) planets with period less than one day have also been detected \citep[e.g.][]{Winn_2018, Livingston_2018, Santerne_2018}. Additionally, these planets seem to have low eccentricities \citep[e.g.,][]{Lithwick_2012, Van_2015}. It is suggested that gravitational interactions between planets in these compact systems helps stabilizing it against perturbations from external bodies. \citep[e.g.][]{Fang_2013, Pu_2015, Volk_2015, Pu_2018, Denham_2018}.

Meanwhile, the population of giant planets and massive stellar companions are also substantial. They are frequently found to inhabit long-period orbits \citep[e.g.,][]{Raghavan_2010, Knutson_2014, Konopacky_2016, Bryan_2016, Zhu_2018, Bryan_2019}, as is the case in our solar system. These faraway and massive companions are crucial in shaping the inner orbits' architecture and, in particular, determining the dynamical stability of the system. For example, a giant planet with an inclined and distant orbit can excite eccentricities and inclinations of the inner planets via the eccentric Kozai-Lidov mechanism through angular momentum exchange \citep[e.g.][]{Takeda_2005, Takeda_2008, Naoz_2016, Denham_2018, Pu_2018}.

The presence of such inclined and massive companions may undermine our ability to detect other planets on the inner orbits \citep[e.g.][]{Hansen_2017, Becker_2017, Mustill_2017}. Specifically, the disrupting effect from the outer companion may change the inclinations of inner orbits \citep[e.g.][]{Becker_2016}, inducing them to oscillate out of the transiting plane, therefore undetectable by the transit method, which is by far the most productive method of detecting exoplanets. Eccentricities of the inner planets may also be excited to such high values that brings about close encounters or orbit crossings, possibly resulting in planets colliding into the star or being ejected out of the system due to the strong gravitational interactions \citep[e.g.][]{Fabrycky_2010, Naoz_2012, Li_2014b, Li_2014Kepler56, Fabrycky_2014, Li_2020}. Consequently, these misaligned, collided, or ejected planets are no longer detectable.

Previous works have studied the secular (i.e., long-term) perturbations both from distant companions and within compact systems. Specifically, the long-time interactions between adjacent planets are studied dating back to Laplace, Lagrange, and Poincar\'e. They found that angular momentum exchange between planets will induce orbital precession, which is later referred to as the Laplace-Lagrange secular theory. The disturbing influence from an inclined distant massive companion is described by the eccentric Kozai-Lidov (EKL) effect \citep[][]{Lidov_1962, Kozai_1962, Naoz_2011, Naoz_2016}. The EKL mechanism dictates that an inclined perturber\footnote{In some cases when both the inner and outer objects are eccentric, a nearly co-planer outer companion can also excite both the eccentricity and inclination of an inner object \citep[see][]{Li_2014a, Li_2014b}. } may excite the eccentricity of an inner planet. However, any effect that causes precession of the inner orbits tends to stabilize the system against the gravitational perturbation from the faraway companion \citep[see][]{Innanen_1997}. Angular momentum exchange between the inner planets turns out to be one of the stabilizing effects \citep[][]{Denham_2018, Pu_2018}. For example, \citet{Denham_2018}, focusing on systems with $2$ planets and a faraway companion ($2+1$ configuration), derived an analytical criterion and tested the stability of the system. The method proposed in this work can be used to estimate and constrain the parameter space of a possible hidden companion of an observed multi-planet system. Furthermore, \citet{Boue_2014} studied the stellar spin-orbit excitations in compact planetary systems with an inclined companion. \citet{Martin_2015} also examined the effect of a distant star on a circumbinary system.

Aside from the planet-planet interactions, the influence of the star on the stability of the system can not be neglected when the orbits of the planets are close enough to the star. Many such close planets are already observed, such as in planetary systems TRAPPIST-1, Kepler-20, Kepler-42, Kepler-90, to name a few. General relativity (GR) also causes precession of the orbits, where due to the weak field nature of the problem we consider only the first-order post-Newtonian approximation. When the summation of the precession rates of GR and Laplace-Lagrange effect is much faster than that of EKL's, further eccentricity excitations will be suppressed. However, when the synthesized precession rate of GR and Laplace-Lagrange effect is comparable to or slower than that of EKL's, a resonant-like behavior takes place which tends to excite the eccentricity of the inner member \citep[e.g.][]{Naoz_2013a}.

In this work, we extend the previous stability criterion derived in \citet{Denham_2018} by incorporating short-range forces or effects, such as general relativity, and comparing the precession rates of different effects. Moreover, we generalize the treatment beyond $2+1$ to general $N+1$ systems and show $2+1$ and $3+1$ systems as examples. We find that general relativity indeed suppresses the excitation from the outer companion and helps stabilize the system. The example systems are numerically integrated to test our analytical criterion. Finally, we apply our criterion to two observed systems to constrain the parameter space in which a hidden companion may reside.

Below we first introduce the interactions and relevant equations in Section~\ref{sec:equations}. Then in Section~\ref{sec:criterion}, we develop a relativistic analytical stability criterion that predicts the long-time stability of a $N+1$ system from its initial conditions by comparing the precession rates of different dynamical effects. Numerical methods used in the rest of this work to test the criterion are introduced in Section~\ref{sec:numerical}. As a proof-of-concept, we apply and test the criterion on a set of hypothetical $2+1$ and $3+1$ systems in Section~\ref{sec:2+1 system} and \ref{sec:3+1 system} respectively. After that, we demonstrate how the criterion can be applied to observed systems to constrain the orbital parameters of a potentially undiscovered companion in Section~\ref{sec:application}. Finally, we conclude with a discussion of this study in Section~\ref{sec:discussion}.


\section{Physical Processes and Equations} \label{sec:equations}
\begin{figure}
	\includegraphics[width=\columnwidth]{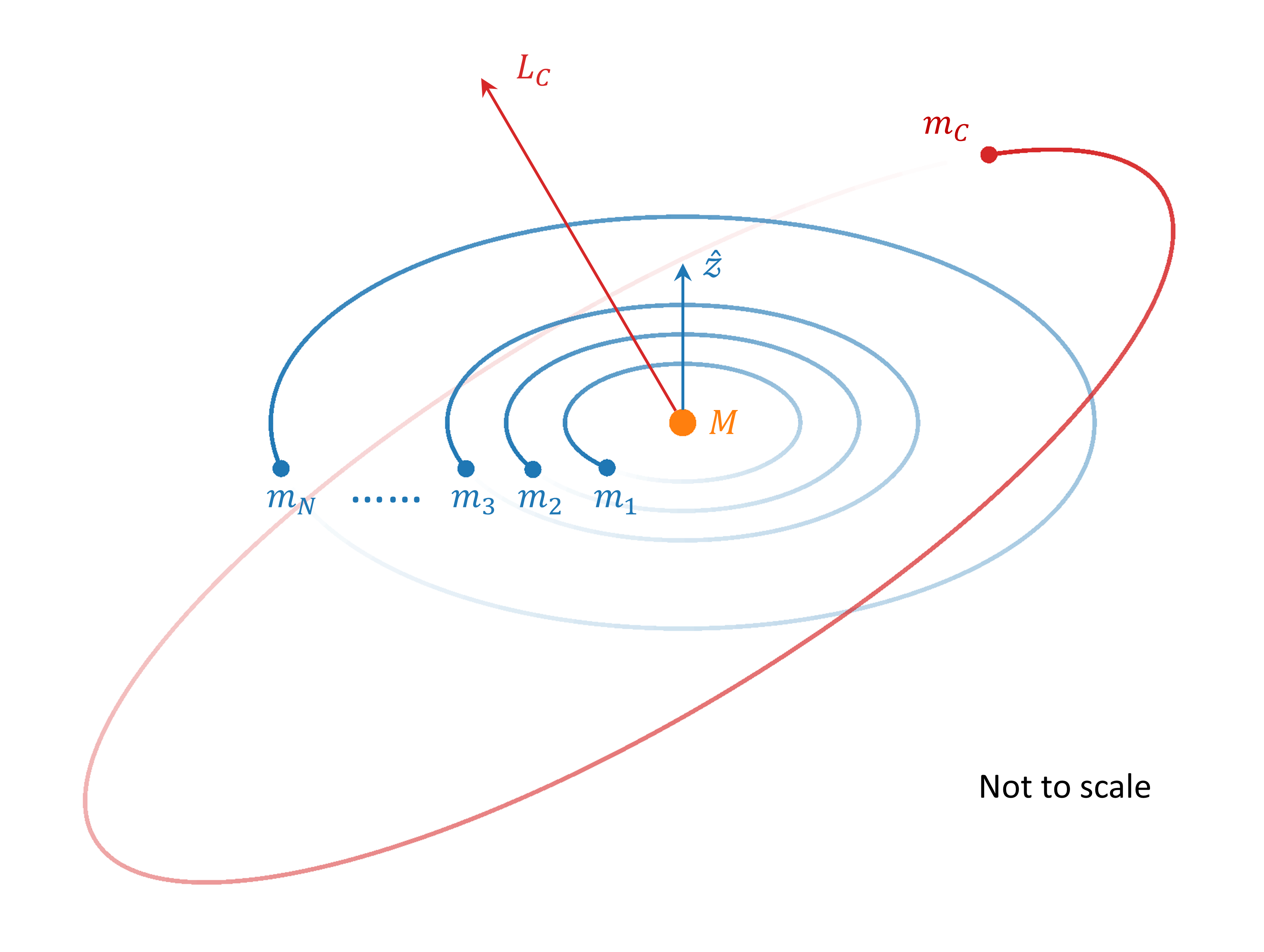}
       \caption{An illustration of the $N+1$ system. We consider $N$ inner planets and a distant companion with masses $m_j$, semi-major axes $a_j$, eccentricities $e_j$, and inclinations $i_j$ where the index $j$ ranges from $1$ to $N$ for the inner planets, and $c$ for the companion. The inner planets are initially on circular and aligned orbits (i.e., small mutual inclinations). The distant companion is misaligned with respect to the inner planets.}
    \label{fig:Cartoon}
\end{figure}

Consider a hierarchical system with $N$ inner planets and a distant companion indexed as $1, 2, ..., N$ for the planets and $c$ for the companion orbiting around a star of mass $M$. Their corresponding masses, semi-major axes, eccentricities, inclinations, longitude of the periapsides, longitude of the ascending nodes, and true anomalies are denoted by $m_j, a_j, e_j, i_j, \varpi_j, \Omega_j$, and $\nu_j$ respectively, where the subscript index $j$ ranges from $1$ to $N$ for the inner planets and $c$ for the companion. (see Figure \ref{fig:Cartoon} for an illustration of the system). Based on the aforementioned observational findings, we assume that the inner planets are initially on tightly packed and nearly circular orbits with small mutual inclinations. We inspect three major mutual interactions within the system that shapes its dynamical features: the companion-planet interaction described by the eccentric Kozai-Lidov (EKL) effect, the planet-planet Laplace-Lagrange (LL) secular evolution, and the star-planet general relativistic (GR) interaction. Here, we describe the effects that the three interactions have on planet $j$ and present their associated timescales.

\subsection{Eccentric Kozai-Lidov Effect}
\label{subsec: EKL}

In a hierarchical system, the Hamiltonian can be decomposed into the summation of Keplerian terms and a coupling term describing the interaction between planet $j$ and the companion \citep[see][]{Naoz_2016}. This term can be expanded in the power series of $a_j/a_c$, which is a small parameter. By expanding the Hamiltonian up to the quadrupole order and with the help of the equations of motion, we find that the $z-$component of the angular momentum of planet $j$, $J_z=\sqrt{1-e_j^2}\cos{\left(i_j - i_c\right)}$, is conserved. Therefore, the eccentricity and inclination of planet $j$ will oscillate in a trade-off manner due to the presence of the companion. When the eccentricity is excited to extreme values, orbit crossing will happen, therefore destabilizing the system. The associated timescale of the EKL effect from the companion on planet $j$ is \citep[][]{Antognini_2015}
\begin{equation}
    T_{j, {\rm EKL}}\sim\frac{16}{15}\frac{a_c^3}{a_j^{3/2}}\sqrt{\frac{M + \sum_{k=1}^N m_k}{Gm_c^2}}\left(1-e_c^2\right)^{3/2}\ ,
	\label{eq:T_EKL}
\end{equation}
where $G$ is the gravitational constant. Note that this is the shortest timescale for eccentricity excitations (i.e., the quadrupole level expansion). If this short timescale is already suppressed, higher values will also be suppressed.

\subsection{Laplace-Lagrange Secular Evolution}
\label{subsec: LL}
The Laplace-Lagrange (LL) secular interaction describes the long-term perturbations between near planets. Consider two inner planets $j$ and $k$. The LL interaction leads to precession in their orbits through angular momentum exchange. The timescale of the periapsis precession of planet $j$ due to the gravitational perturbation of planet $k$ is given by \citep[e.g.,][Section 7.7]{Murray_2000}
\begin{align}
    T_{j, {\rm LL}} = \Bigg[ & A_{jj} + \sum_{k=1, k\neq j}^{N} A_{jk}\left(\frac{e_k}{e_j}\right)\cos\left(\varpi_k-\varpi_j\right) \nonumber \\
    - & B_{jj} - \sum_{k=1, k\neq j}^{N} B_{jk}\left(\frac{i_k}{i_j}\right)\cos\left(\Omega_k - \Omega_j\right)\Bigg]^{-1} \ ,
    \label{eq:T_LL}
\end{align}
where
\begin{align}\label{eq:A}
    &A_{jj} = + \frac{n_j}{4\pi}\sum_{k=1, k\neq j}^{N} \frac{m_k}{M+m_j}\alpha_{jk}\bar\alpha_{jk} f_\psi(\alpha_{jk}) \ ,\\
    &A_{jk} = - \frac{n_j}{4\pi}\frac{m_k}{M+m_j}\alpha_{jk}\bar\alpha_{jk}f_{2\psi}(\alpha_{jk}) \ ,\\
    &B_{jj} = - \frac{n_j}{4\pi}\sum_{k=1, k\neq j}^{N} \frac{m_k}{M+m_j}\alpha_{jk}\bar\alpha_{jk}f_\psi(\alpha_{jk}) \ ,\\
    &B_{jk} = + \frac{n_j}{4\pi}\frac{m_k}{M+m_j}\alpha_{jk}\bar\alpha_{jk}f_\psi(\alpha_{jk}) \ ,\label{eq:B}
\end{align}
and
\begin{align}
    f_\psi(\alpha_{jk}) = \int_0^{2\pi} \frac{\cos\psi}{{\left(\alpha_{jk}^2 - 2\alpha_{jk}\cos\psi + 1\right)}^{3/2}} d\psi\ , \label{eq:fpsi}\\
    f_{2\psi}(\alpha_{jk}) = \int_0^{2\pi} \frac{\cos2\psi}{{\left(\alpha_{jk}^2 - 2\alpha_{jk}\cos\psi + 1\right)}^{3/2}} d\psi \ . \label{eq:f2psi}
\end{align}
in which $\alpha_{jk}$ and $\bar\alpha_{jk}$ are defined as
\begin{equation}
    \alpha_{jk} = 
    \begin{cases}
        a_j/a_k, & \text{if $a_j < a_k$}\ ,\\
        a_k/a_j, & \text{if $a_j > a_k$}\ ,
    \end{cases}
\end{equation}
and
\begin{equation}
    \bar\alpha_{jk} = 
    \begin{cases}
        a_j/a_k, &  \text{if $a_j < a_k$}\ ,\\
        1, &        \text{if $a_j > a_k$}\ .
    \end{cases}
    \label{eq:bar_alpha}
\end{equation}
These equations were used in \citet{Denham_2018} to find the analytical stability criterion\footnote{Note that  \citet{Denham_2018} had a typo in the equivalent equations for the $A$ and $B$ coefficients (Equation~(\ref{eq:A})$-$(\ref{eq:B})), but it did not propagate into the actual calculations.} in the absence of general relativity effects. Here the equations are generalized to account for the influence on planet $j$ from more than one planets.

Substituting Equation~(\ref{eq:A}) $-$ (\ref{eq:bar_alpha}) into Equation~(\ref{eq:T_LL}) yield
\begin{align}
\label{eq:T_LL combined}
    T_{j, {\rm LL}} &= \frac{4\pi}{n_j}\Bigg\{\sum_{k=1, k\neq j}^{N} \frac{m_k}{M+m_j}\alpha_{jk}\bar\alpha_{jk} \nonumber \\
    & \times \bigg[\left(2-\frac{i_k}{i_j}\cos{\left(\Omega_k - \Omega_j\right)}\right) f_\psi(\alpha_{jk}) \nonumber \\
    & - \frac{e_k}{e_j}\cos{\left(\varpi_k - \varpi_j\right)} f_{2\psi}(\alpha_{jk})\bigg]\Bigg\}^{-1}\ .
\end{align}
In addition to \citet{Denham_2018}, we find that during relativistic evolution, $\cos(\varpi_k-\varpi_j)$ and $\cos(\Omega_k-\Omega_j)$ in Equation~(\ref{eq:T_LL combined}) can both vary between $-1$ and $1$. Combined with the fact that the integral in Equation~(\ref{eq:fpsi}) and Equation (\ref{eq:f2psi}) are positive for any given $\alpha_{jk}$ between $0$ and $1$, we can therefore constrain the Laplace-Lagrange precession timescale with an upper and a lower limit by allowing the cosine terms to vary to their limits:
\begin{align}
    T_{j, {\rm LL}, \min} = & \frac{4\pi}{n_j}\Bigg\{\sum_{k=1, k\neq j}^{N} \frac{m_k}{M+m_j}\alpha_{jk}\bar\alpha_{jk} \nonumber\\
    & \times \left[\left(2 + \frac{i_k}{i_j}\right) f_\psi(\alpha_{jk}) + \frac{e_k}{e_j} f_{2\psi}(\alpha_{jk})\right]\Bigg\}^{-1}\ , \label{eq:T_LL min} \\
    T_{j, {\rm LL}, \max} = & \frac{4\pi}{n_j}\Bigg\{\sum_{k=1, k\neq j}^{N} \frac{m_k}{M+m_j}\alpha_{jk}\bar\alpha_{jk} \nonumber\\
    & \times \left[\left(2 - \frac{i_k}{i_j}\right) f_\psi(\alpha_{jk}) - \frac{e_k}{e_j} f_{2\psi}(\alpha_{jk})\right]\Bigg\}^{-1}\ . \label{eq:T_LL max}
\end{align}
We will soon show in Section~\ref{sec:criterion} that the two limits enclose an area in the parameter space, which is a transition zone between stable and unstable systems. If the change in the orientation of the planet’s orbit is fast enough to average out the eccentricity excitations brought by the eccentric Kozai-Lidov effect, the system will remain stable over time \citep[e.g.,][]{Denham_2018,Pu_2018}.

\subsection{General Relativity}
\label{subsec: GR}
General relativity is another effect that induces orbital precession for close-in planets. When relativistic precession is fast enough, the EKL eccentricity excitation can be suppressed. The associated timescale of planet $j$ is the period for its orbit to precess one cycle \citep[e.g.,][]{Misner_1973}:
\begin{equation}
    T_{j, {\rm GR}} = 2\pi\frac{c^2a_j^{5/2}(1-e_j^2)}{3\left(GM\right)^{3/2}} \ .
    \label{eq:T_GR}
\end{equation}

The shortest eccentricity excitation timescale due to the perturbations from the companion is proportional to the EKL-quadrupole timescale (i.e., Equation~(\ref{eq:T_EKL})). Therefore, the maximum eccentricity that planet $j$ can be excited to is due to the  quadrupole level of the EKL approximation. For an initially circular inner orbit, the maximum eccentricity that planet $j$ can reach throughout its evolution in the presence of GR is determined by the cubic equation
\begin{equation}
    \epsilon_{\rm GR} = \frac{9}{8} \frac{J_{\rm min}+1}{J_{\rm min}}\left(J_{\rm min}^2 - \frac{5}{3}\cos^2{i_c}\right)\ ,
    \label{eq:emax}
\end{equation}
where $J_{\rm min}=\sqrt{1-e_{j, {\rm max}}^2}$ is the minimum dimensionless angular moment and $\epsilon_{\rm GR}=\left(1-e_j^2\right)T_{\rm EKL}/T_{\rm GR}$  \citep[][]{Liu_2015, Naoz_2016}. Notice that $\epsilon_{\rm GR}\sim1$ when $T_{\rm GR}$ and $T_{\rm EKL}$ are of comparable magnitude, and $e_j \approx 0$ in our configuration. It is worth mentioning that if the inner orbit is not initially circular then the resonance nature changes and the maximum eccentricity takes a different value \citep[e.g.,][]{Hamers_2020, Hansen_2020}. 
Solving for $e_{j, {\rm max}}$ via Equation~(\ref{eq:emax}) and substituting it back in Equation~(\ref{eq:T_GR}) gives the shortest timescale of relativistic precession $T_{\rm GR, \min}$.
\begin{equation}
    T_{j, {\rm GR, min}} = 2\pi\frac{c^2a_j^{5/2}(1-e_{j, {\rm max}}^2)}{3\left(GM\right)^{3/2}} \ .
    \label{eq:T_GR_min}
\end{equation}

In what follows we generalize the stability criterion developed in \citet{Denham_2018} to (1) include GR effects, (2) expand beyond two planets with a companion (i.e., $2+1$) to multi-planets with a companion (i.e., $N+1$) configuration.


\section{Analytical Stability Criterion}
\label{sec:criterion}

As discussed in Section~\ref{sec:equations}, the distant companion can excite the eccentricities and inclinations of the inner orbits. Meanwhile, GR along with other dynamical processes, such as Laplace-Lagrange and EKL, induces precession of the inner orbits. If the combined precession rate induced by GR and Laplace-Lagrange effect is faster than that of EKL, eccentricity excitations will be suppressed \citep[e.g.,][]{Naoz_2013a, Denham_2018}\footnote{Note that in some cases the Laplace-Lagrange level of interaction can excite eccentricities between the inner orbits, but GR precession may also suppress these excitations \citep[e.g., Section 7.7][]{Murray_2000, Faridani_2021}}. Therefore, to explore whether a system is stable, we need to inspect and compare all the precession rates among the aforementioned processes (see Section \ref{sec:equations}). 

The inner system undergoes accumulated precession due to GR, Laplace-Lagrange, and any other short-range effects that may affect the system, such as tidal and magnetic interactions \citep[e.g.,][]{Hut_1981, Eggleton_1998, Eggleton_2001, Fabrycky_2007, Liu_2015, Ahuir_2021}. In fact, any short-range effect that causes apsidal precession can suppress the eccentricity excitations brought by the EKL effect \citep[e.g.][]{Innanen_1997, Liu_2015}. For the inner planets to survive, these precession rates need to be faster than the induced precession from the EKL, which is responsible for the eccentricity excitations. 
Therefore, the general stability criterion can be written as:
\begin{equation}
    \label{eq:criterion}
    \dot \omega_{\rm EKL}
    \begin{cases}
        < \dot \omega_{\rm LL} + \dot \omega_{{\rm GR}, \max} + \dot \omega_{\rm SR}, \quad \textrm{stable}\ ,\\
        > \dot \omega_{\rm LL} + \dot \omega_{{\rm GR} , \max}  + \dot \omega_{\rm SR}, \quad \textrm{unstable} \ ,
    \end{cases}
\end{equation}
where $\dot \omega_{\rm LL} = 1/T_{\rm LL}$, $\dot \omega_{{\rm GR} , \max}=1/T_{{\rm GR}, \min}$, $\dot \omega_{\rm EKL} = 1/T_{\rm EKL}$, and $\dot \omega_{\rm SR}$ stands for the precession rate of other short-range forces or effects. Notice that while in the previous section we considered the precession timescale, we find that the stability criterion is better characterized in terms of the time derivative of the resonant angle, i.e., the precession rate of the argument of periapsis. Even though it is not the resonant angle of the Laplace-Lagrange secular evolution, the argument of periapsis is relevant because it is the resonant angle of the EKL quadrupole level of approximation. Moreover, the precession rates of different stabilizing effects can be added directly because the argument of periapsis, $\omega$, is one of the canonical variables of the system, the Delaunay's elements \citep[e.g.,][]{Valtonen_2006}. According to the Hamilton's equations, the time derivative of $\omega$ can be added because the Hamiltonian of interactions between different objects can be added directly. 

Equation~(\ref{eq:criterion}) generalizes the previous analytical criterion which neglected other short-range effects and is straightforward to extend beyond $2+1$ configuration to multi-planet systems. We first introduce the numerical methods we use throughout this work in Section \ref{sec:numerical}. In Section~\ref{sec:2+1 system} we then consider a two-planet system with one far-away companion ($2+1$), and extend the stability criterion to include GR effects. Furthermore, in Section~\ref{sec:3+1 system} we show the extension of the criterion to $3+1$ systems.


\section{Numerical Methods}
\label{sec:numerical}

In the rest of this work, we test our analytical criterion with numerical approaches. We use a combination of the rings code\footnote{Available at \url{https://github.com/farr/Rings}.} and the N-Body integration package REBOUNDx \citep[][]{rebound, reboundx}.

The rings code adopts the Gauss's averaging method, which considers the interactions between planets by treating each object as a ring of material whose mass is spread across its orbit, instead of a mass point \citep[][Section 7.6]{Murray_2000}. Orbital evolution is then calculated by the interactions between different rings. The rings code is an adaptive time-step integrator, and the initial time step is not specified and left as default in our simulations. The robustness of this algorithm was verified in many different studies for non-crossing orbits \citep[e.g.,][]{Touma_2009, Touma_2014, Nesvold_2016,  Sridhar_2016, Denham_2018}. We modified the rings code to include the GR effect up to the first order post-Newtonian expansion
\citep[e.g.,][]{Tremaine_2012, Naoz_2013a, Will_2017}\footnote{Available at \url{https://github.com/wei-lingfeng/Rings-GR}.}. The Gauss's averaging method provides a relatively fast integration for systems that do not experience close-encounters. Therefore, we primarily use this method for a few slowly-varying stable systems.

REBOUNDx is a direct N-body simulation code with GR effects we primarily use. We choose the ``WHFast'' integrator \citep[][]{wh, reboundwhfast} with the default time-step of $0.001/2\pi$ year in all of our simulations. We repeatedly integrate systems on a basis of $24$ hours of running time and inspect the results at the end of each day's run. This corresponds to roughly $3-6$ Myr of system evolution time every $24$ hours depending on different configurations. Systems are integrated until the number of unstable systems remains unchanged between two consecutive 24-hour runs. Upper limits of system evolution time are given case-by-case in corresponding sections below. N-body integration with GR is significantly slower but more accurate for close-encounters \citep[see][]{reboundx}. Therefore, all the unstable systems are integrated using this method. We double-checked the consistency between the rings code and REBOUNDx by comparing their simulation results on the same systems and found conformity between them.

As \citet{Denham_2018} did, we utilize the same proxy $log(1-\delta)$ to quantify the stability of the system, where $\delta$ is a parameter representing the closest distance between two inner orbits of planet $j$ and $k$ throughout their evolution:
\begin{equation}
    \delta \coloneqq \min\left[\frac{a_j(1-e_j)-a_k(1+e_k)}{a_j-a_k}\right]\ ,
    \label{eq:delta}
\end{equation}
assuming $j>k$. The closer the inner orbits are, the smaller $\delta$ is. When the two orbits cross, i.e., $\log(1-\delta) \ge 0$, strong gravitational interactions will almost certainly result in unpredictable and violent behaviors. Therefore, systems that undergo orbit crossings are defined as “unstable” systems.

We are aware that there are different proxies which can be used to indicate whether a system is stable or not, such as Lyapunov characteristic exponent and the Mean Exponential Growth of Nearby Orbits (MEGNO, see \citet{Cincotta_2000, Gozdziewski_2001}). However, instead of inspecting the rate of separation of different orbits, focusing only on the separation itself serves its purpose in this work, as we will show in Section \ref{sec:2+1 system} and \ref{sec:3+1 system}. Therefore, we adopt the more easily calculated and more intuitive proxy $\delta$ defined in Equation~(\ref{eq:delta}).


\section{Example of 2+1 Systems}
\label{sec:2+1 system}

As a proof-of-concept, we consider two sub-Earth-sized inner planets and a distant Jupiter-sized companion orbiting around a $1M_\odot$ star. We initialize the two inner planets on nearly circular orbits with low mutual inclinations. Planet $1$'s semi-major axis is $0.03$~au and planet $2$'s semi-major axis is allowed to vary from $0.05-0.3$~au. The outer companion is initialized on a $5$-au orbit with an inclination of $85^\circ$ with respect to the inner ones\footnote{The value of the initial mutual inclination has a negligible effect on the stability, as long as it is higher than the Kozai angle \citep{Denham_2018}. }. The eccentricity of the companion is chosen as another free parameter that ranges from $0-0.9$. The initial conditions are listed in Table~\ref{tab:2+1 initial conditions}. All other orbital elements that are not listed here are initialized to be 0, including the argument of periapsis $\omega_j$, the longitude of ascending node $\Omega_j$, and the true anomaly $\nu_j$.

\begin{table}
	\centering
	\caption{Initial conditions of 2+1 systems. The $2+1$ configuration is initialized with two sub-Earth sized inner planets and a massive, distant companion. The semi-major axis of planet $2$ and the eccentricity of the companion are chosen as two free parameters that varies within a specific range. Eccentricities and inclinations of inner planets are set to small values to avoid peculiar behaviors of the integrator.}
	\label{tab:2+1 initial conditions}
	\begin{tabular}{lcccc} 
		\hline
		Object & Mass ($M_\odot)$ & $a$ (au) & $e$ & $i$\\
		\hline
		Star & $1$ & & & \\
		Planet $1$ & $10^{-6}$ & $0.03$ & $10^{-3}$ & $10^{-3}$ rad\\
		Planet $2$ & $10^{-6}$ & $0.05 - 0.3$ & $10^{-3}$ & $10^{-3}$ rad\\
		Companion & $10^{-3}$ & $5$ & $0 - 0.9$ & $85^{\circ}$\\
		\hline
	\end{tabular}
\end{table}

\begin{figure}
    \centering
    \includegraphics[width=0.5\textwidth]{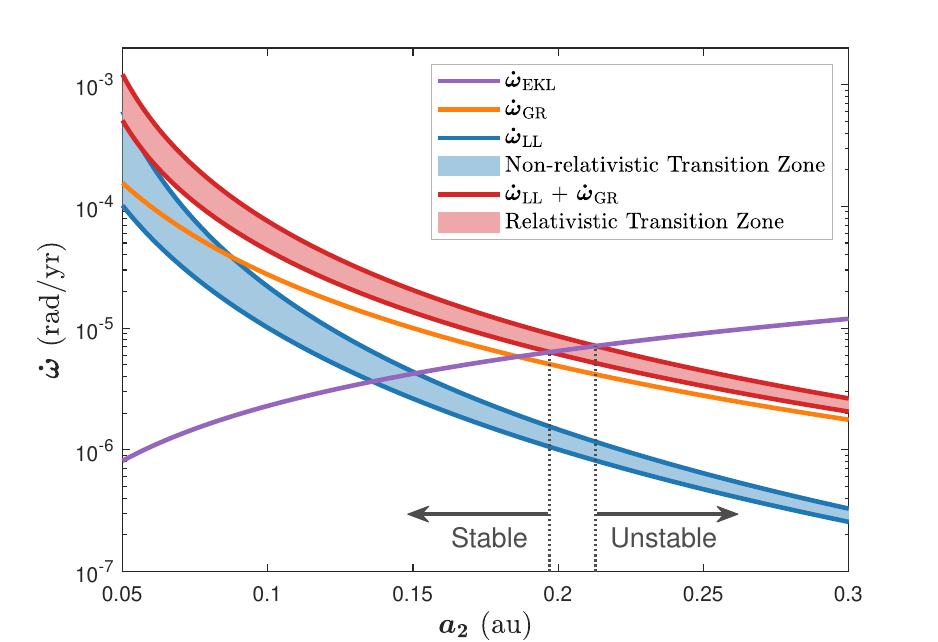}

    \caption{Precession rates of different effects on planet $2$. We choose $e_c=0.5$ to show the dependency of precession rates of various effects on $a_2$. When $\dot \omega_{\rm LL} + \dot \omega_{\rm GR}$ (red curve) is faster than $\dot \omega_{\rm EKL}$ (purple curve), the system is predicted stable, and unstable vice versa. Within the transition zone, the system are expected to undergo large eccentricity excitations, but orbit-crossing may happen late in time.}
    \label{fig:precession_2+1}
\end{figure}

In this specific $2+1$ configuration, we find that within the varying range of the free parameters in Table \ref{tab:2+1 initial conditions}, the corresponding GR precession rates of planet $1$ is always much faster compared to EKL's and LL's, which is reasonable as it is the closest planet to the star and the furthest one from the disturbing companion. Consequently, planet $1$ is always stable in our configuration. Therefore, we focus on the precession rates of planet $2$ from now on.

The precession rate of each effect on planet $2$ with the choice of $e_c=0.5$ as an example is plotted as a function of $a_2$ in Figure~\ref{fig:precession_2+1}. As can be seen, the precession rate of GR is generally faster than that of LL in this case. Consequently, the intersections between $\dot \omega_{\rm LL} + \dot \omega_{\rm GR}$ (red curves) and $\dot \omega_{\rm EKL}$ (blue curves), or the maximum $a_2$ that planet $2$ can survive for a long time, is relaxed to higher values compared with non-relativistic predictions.

\begin{figure*}
    \centering
    \includegraphics[width = 0.8\textwidth]{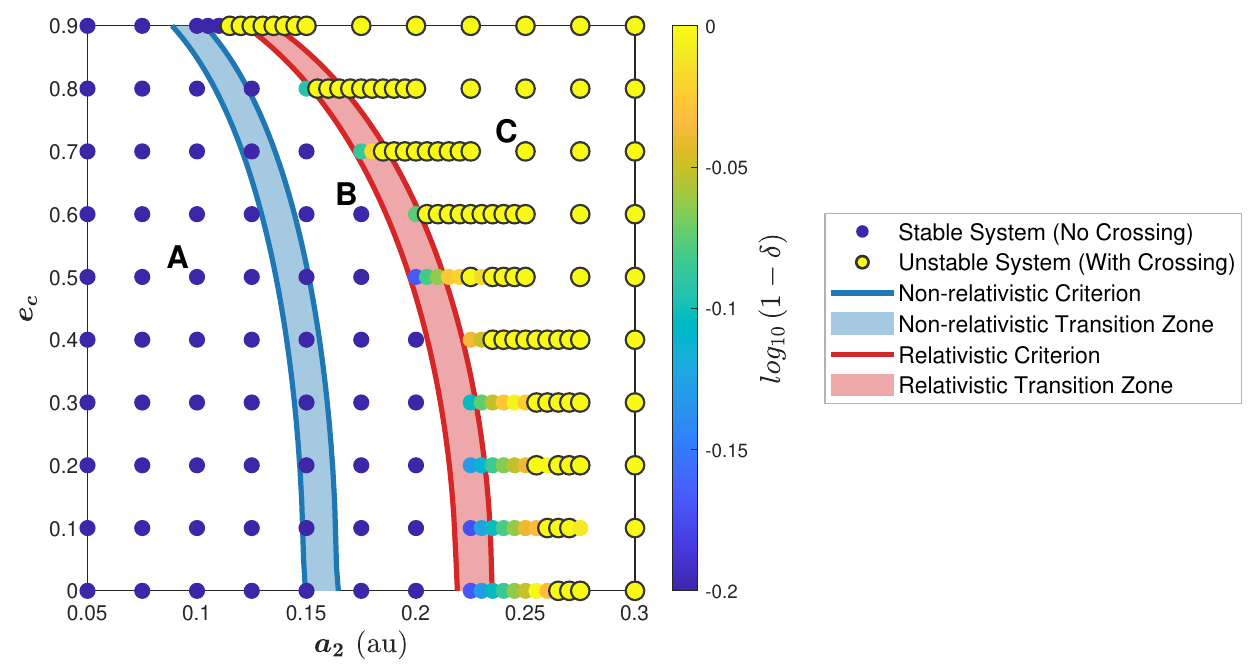}
    \caption{Stability criterion and simulation results for $2+1$ systems. We compare the stability criterion with simulation results for $2+1$ systems. Relativistic and non-relativistic stability criterion based on Equation~(\ref{eq:e_crit 2+1}) are plotted in red and blue curves respectively. Each scattered point represents a numerically integrated system accounting for GR effects in the $e_c-a_2$ parameter space. Systems are integrated up to a maximum of $170$~Myr. The color code represents how close the orbit of the two inner planets become defined by $\log (1-\delta)$ (see Equation~(\ref{eq:delta})) throughout their evolution. Deep blue represents large separations between planet $1$ and planet $2$ without orbit crossing, whereas yellow stands for close encounters. The points with the black contours represent systems that experience orbit crossing. Here we plot all the values of $\log (1-\delta)$ at or below $-0.2$ in the same deepest blue color in order to show the transition from stable to unstable systems more clearly. In Figure \ref{fig:Evolution_2+1} we show the time evolution of three representative systems in different regions marked as A, B, and C in this plot.}
    \label{fig:Delta_2+1}
\end{figure*}

\begin{figure*}
    \centering
    \includegraphics[width = \textwidth]{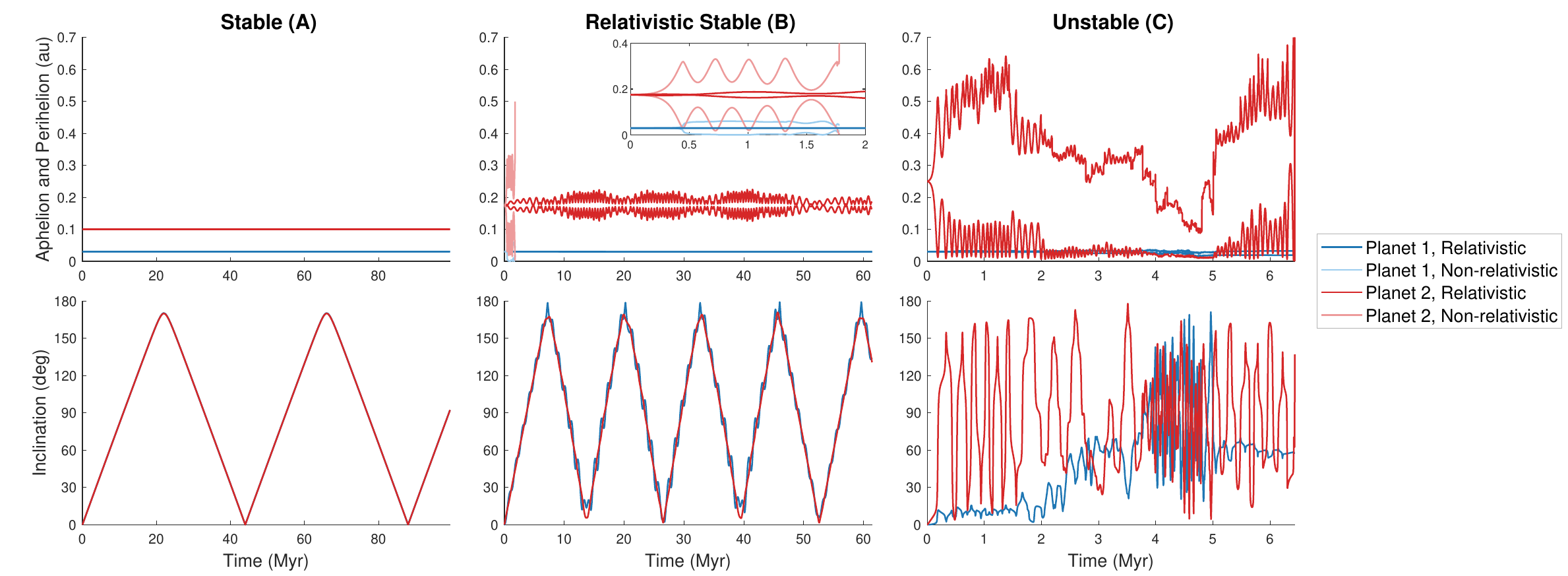}
    \caption{Time evolution of three representative systems. We show the time evolution of aphelion and perihelion (top row) and inclination (bottom row) for planet $1$ (solid blue) and planet $2$ (solid red) in three representative relativistic systems labeled as A, B, and C in Figure~\ref{fig:Delta_2+1}. For system B, non-relativistic evolution of aphelion and perihelion are overplotted in light blue and light red for comparison. Inset figure in the top-middle panel zoom in the first $2$ Myr to illustrate the system behavior in the non-relativistic case. The three systems share the following initial orbital parameters in common: $M=1M_\odot$, $m_1=m_2=10^{-6}M_\odot$, $m_c=10^{-3}M_\odot$, $a_1=0.03$ au, $a_c=5$ au, $e_1=e_2=0.001$, $i_1=i_2=0.001$ rad, $i_c=85^\circ$, and $\omega_1=\omega_2=\omega_c=\Omega_1=\Omega_2=\Omega_c=0$. The only differences are $a_2$ and $e_c$. For the stable system A, $a_2=0.1$, $e_c=0.5$. For relativistic stable system B (only stable when considering GR), $a_2=0.175$, $e_c=0.6$. Finally, for unstable system C, $a_2=0.25$, $e_c=0.7$.}

    \label{fig:Evolution_2+1}
\end{figure*}

From the timescale argument presented in Section~\ref{sec:equations}, we can find a stability criterion for the companion's eccentricity as a function of planet $2$'s semi-major axis $a_2$ according to Equation~(\ref{eq:criterion}). The stability criterion can be easily rearranged to account for other unknown variables in the system (for example, companion eccentricity and period) for an observed system see Section~\ref{sec:application}.
Equating the two sides in Equation~(\ref{eq:criterion}) with the choice of $j=2$ yields a critical eccentricity for the companion:
\begin{equation}
    e_{c, {\rm crit}} = \left[1-\left(\frac{15}{16} \frac{a_2^{3/2}}{a_c^3} \sqrt{\frac{G}{M+\sum_{k=1}^2 m_k}} \frac{m_c}{\dot \omega_{2, {\rm stab}}}\right)^{2/3}\right]^{1/2} \ ,  \\
    \label{eq:e_crit 2+1}
\end{equation}
where $\dot\omega_{2, {\rm stab}}$ stands for the summation of all the precession rates of stabilizing effects on planet $2$:
\begin{equation}
    \dot \omega_{2, {\rm stab}} =  \dot \omega_{2, {\rm LL}} + \dot \omega_{2, {\rm GR}, \max} \ .
\end{equation}
In Equation~(\ref{eq:e_crit 2+1}), $e_{c,{\rm crit}}$ denotes the critical eccentricity of the companion that separates the stable and unstable systems in the $e_c - a_2$ parameter space. Note that the limits of the critical eccentricity, $e_{c, {\rm crit}}(\dot \omega_{2, {\rm LL}, \min})$ and $e_{c, {\rm crit}}(\dot \omega_{2, {\rm LL}, \max})$, are obtained by substituting the minimum and maximum $\dot \omega_{\rm LL}$ (reciprocal of Equation~(\ref{eq:T_LL min}) $-$ (\ref{eq:T_LL max})) into Equation~(\ref{eq:e_crit 2+1}). Our analytical criterion in Equation~(\ref{eq:criterion}) then becomes
\begin{equation}
    e_c
    \begin{cases}
        < e_{c, {\rm crit}}(a_2, \dot \omega_{2, {\rm LL}, \min}), \quad \textrm{stable}\ ,\\
        > e_{c, {\rm crit}}(a_2, \dot \omega_{2, {\rm LL}, \max}), \quad \textrm{unstable}\ .
    \end{cases}
    \label{eq:criterion 2+1}
\end{equation}
The region between the two critical eccentricities is a transition zone from stable to unstable systems, where systems are nearly or completely unstable. It is straightforward to incorporate other short-range physical processes in addition to GR by simply adding corresponding terms to $\dot \omega_{2, {\rm stab}}$.

To test the criterion derived above in a numerical approach, we integrate the $2+1$ systems with planet $2$'s semi-major axis increasing from $0.05$ to $0.3$~au in steps of $0.025$~au, and the eccentricity of the distant companion ranges from $0$ to $0.9$ in steps of $0.1$. In addition, we add more densely populated integrated systems in the parameter space where it is close to the critical eccentricity $e_{c, {\rm crit}}$  (see Figure~\ref{fig:Delta_2+1}). The systems are integrated up to $170$~Myr or until orbit crossing. As mentioned, we use a combination of Gauss's method and N-body integration to check for consistency. 

In Figure~\ref{fig:Delta_2+1}, each scattered point represents a simulated system, and their colors reflect the minimum distance between planet $1$ and $2$ ($\delta$ in Equation~(\ref{eq:delta})) for each system throughout its evolution. Stable orbits are represented as deep blue colors, whereas closer or even crossing orbits are plotted in yellow. All the systems that experience orbit-crossing are circled out with a black contour. Here we impose a minimum cutoff value of $-0.2$ on $\log_{10}\left(1-\delta\right)$ in color representation. In other words, all stable systems whose $\log_{10}\left(1-\delta\right)$ are less than $-0.2$ are represented as the same deepest blue color, so as to better illustrate the transition from stability to instability. For comparison, the relativistic and non-relativistic analytical stability criteria derived in the presence of GR and in \citet{Denham_2018} are over-plotted in red and blue curves respectively. Relativistic systems are predicted to be stable below the red curves and unstable above them. Shaded regions represent their corresponding transition zones, in which systems are likely to experience large eccentricity oscillations but orbit crossing happens after a long time, possibly exceeding the limit of our integration time. Systems between the blue and red curves are what relativistic and non-relativistic criterion predicts differently. Compared with the non-relativistic criterion, the relativistic one expands the stable region in the parameter space. 

There is a slight discrepancy on the lower-right part of the parameter space due to the limited integration time. We emphasize that the criterion developed here predicts unstable systems if orbit crossing happens at any time during their evolution regardless of timescale. Therefore, the simulated results may not reach instability within the limited numerical integration time. Integrating to Gyrs is challenging in the presence of GR precession. Thus, the simulation results tend to underestimate the unstable region in the parameter space \citep[e.g.,][]{Myllari_2018}. Nonetheless, the systems that are expected unstable but do not experience orbit crossing all have very close orbital distance $\delta$ (as close as $10^{-3}$~au). As depicted in Figure~\ref{fig:Delta_2+1}, our analytical prediction is close to the numerical results.  

To illustrate the effect of the GR precession in stabilizing the system, we select three representative systems from the grid, marked as A, B, and C as indicated in Figure~\ref{fig:Delta_2+1}. They correspond to stable, relativistic stable (unstable by non-relativistic predictions) and unstable systems, respectively. The evolution of the aphelion, perihelion, and inclination are plotted as a function of time in Figure~\ref{fig:Evolution_2+1}. 

According to the non-relativistic criterion, system B should be unstable. Indeed, when we integrate system B without adding GR effects, the orbits cross and planet $2$ collides with the star after about $1.5$ million years, as can be seen from the zoomed-in plot in the top middle panel of Figure~\ref{fig:Evolution_2+1}. However, with the help of general relativity, the excited oscillations in eccentricity and inclination from the distant companion are largely suppressed, rendering stable evolution. The simulation results validates the relativistic analytical criterion proposed in Section~\ref{sec:criterion}.


\section{Example of 3+1 Systems}
\label{sec:3+1 system}

As another example of the criterion's application, we now inspect the stability of systems with three inner planets and a distant companion, or $3+1$ systems. The systems are initialized with three close-in sub-Earth-sized planets and a Jupiter-sized companion orbiting around a star of $M=1M_\odot$. Planet $1$ and the companion lie on the same orbit as in the $2+1$ case. The initial conditions of the system are listed in Table \ref{tab:3+1 initial conditions}.

\begin{table}
	\centering
	\caption{Initial conditions for $3+1$ systems. All other unspecified orbital elements are initialized to be zero.}
	\label{tab:3+1 initial conditions}
	\begin{tabular}{lcccc} 
		\hline
		Object & Mass ($M_\odot$) & $a$ (au) & $e$ & $i$\\
		\hline
		Star & $1$ & & & \\
		Planet $1$ & $10^{-6}$ & $0.03$ & $10^{-3}$ & $10^{-3}$ rad\\
		Planet $2$ & $10^{-6}$ & $0.05$ & $10^{-3}$ & $10^{-3}$ rad\\
		Planet $3$ & $10^{-6}$ & $0.1 - 0.3$ & $10^{-3}$ & $10^{-3}$ rad\\
		Companion & $10^{-3}$ & $5$ & $0 - 0.9$ & $85^{\circ}$\\
		\hline
	\end{tabular}
\end{table}

\begin{figure}
    \centering
    \includegraphics[width=\columnwidth]{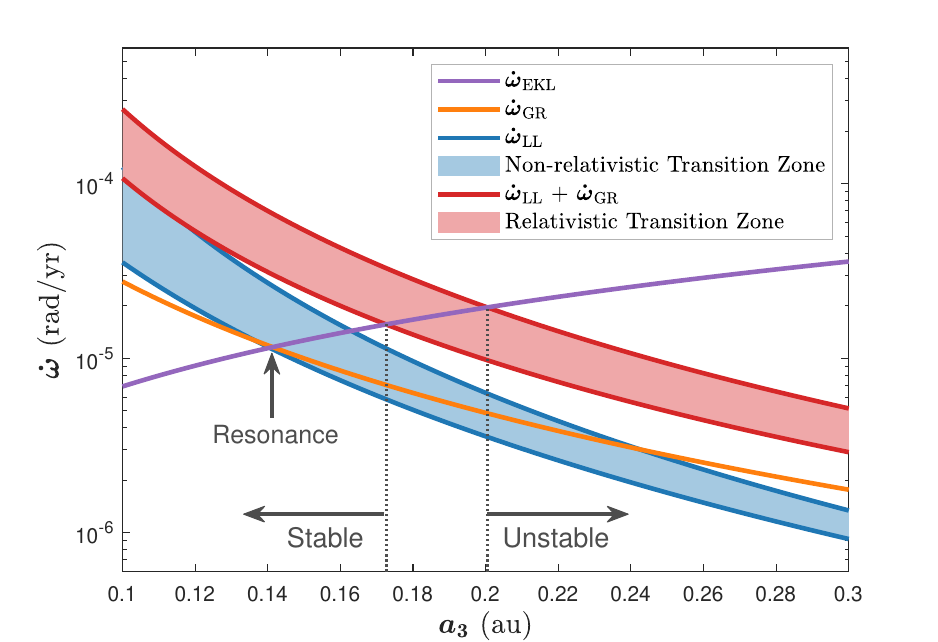}
    \caption{Precession rates of different effects on planet $3$. We choose $e_c=0.8$ to show the dependency of precession rates on $a_3$. When $\dot \omega_{\rm LL} + \dot \omega_{\rm GR}$ (red curve) dominates over $\dot \omega_{\rm EKL}$ (purple curve), the system is predicted stable, and unstable vice versa. Within the transition zone, the system is expected to experience large eccentricity excitations, but orbit-crossing may happen late in time. When the precession rates from the three effects are equal, we observe a resonance-like behavior in the simulation, which is discussed in Appendix~\ref{appenix} in detail.}
    \label{fig:precession_3+1}
\end{figure}

After applying the generalized stability criterion Equation~(\ref{eq:criterion}) to each planet, we find that planet $1$ and $2$ are stable in the varying range of $a_3$ and $e_c$. However, planet $3$ is not. The precession rates of different effects on planet $3$ with the choice of $e_c=0.8$ as an example is plotted in Figure~\ref{fig:precession_3+1}. 

As before, equating the two sides in Equation~(\ref{eq:criterion}) for planet $3$ gives a similar critical eccentricity of the companion $e_{c, {\rm crit}}$:
\begin{equation}
    e_{c, {\rm crit}} = \left[1-\left(\frac{15}{16} \frac{a_3^{3/2}}{a_c^3} \sqrt{\frac{G}{M + \sum_{k=1}^3 m_k}} \frac{m_c}{\dot \omega_{3, {\rm stab}}}\right)^{2/3}\right]^{1/2} \ ,
    \label{eq:e_crit 3+1}
\end{equation}
where
\begin{equation}
    \dot \omega_{3, {\rm stab}} =  \dot \omega_{3, {\rm LL}} + \dot \omega_{3, {\rm GR}, \max} \ .
\end{equation}
The general criterion in Equation~(\ref{eq:criterion}) then becomes
\begin{equation}
    e_c
    \begin{cases}
        < e_{c, {\rm crit}}(a_3, \dot \omega_{3, {\rm LL}, \min}), \quad \textrm{stable}\ ,\\
        > e_{c, {\rm crit}}(a_3, \dot \omega_{3, {\rm LL}, \max}), \quad \textrm{unstable}\ .
    \end{cases}
    \label{eq:criterion 3+1}
\end{equation}
The only difference from the $2+1$ case is the subscripts changing from $2$ to $3$. Recall that $\dot \omega_{3, {\rm LL}}$ includes the contribution of Laplace-Lagrange effects from all the planets, according to Equation~(\ref{eq:T_LL min}) and (\ref{eq:T_LL max}).

\begin{figure*}
    \centering
    \includegraphics[width = 0.8\textwidth]{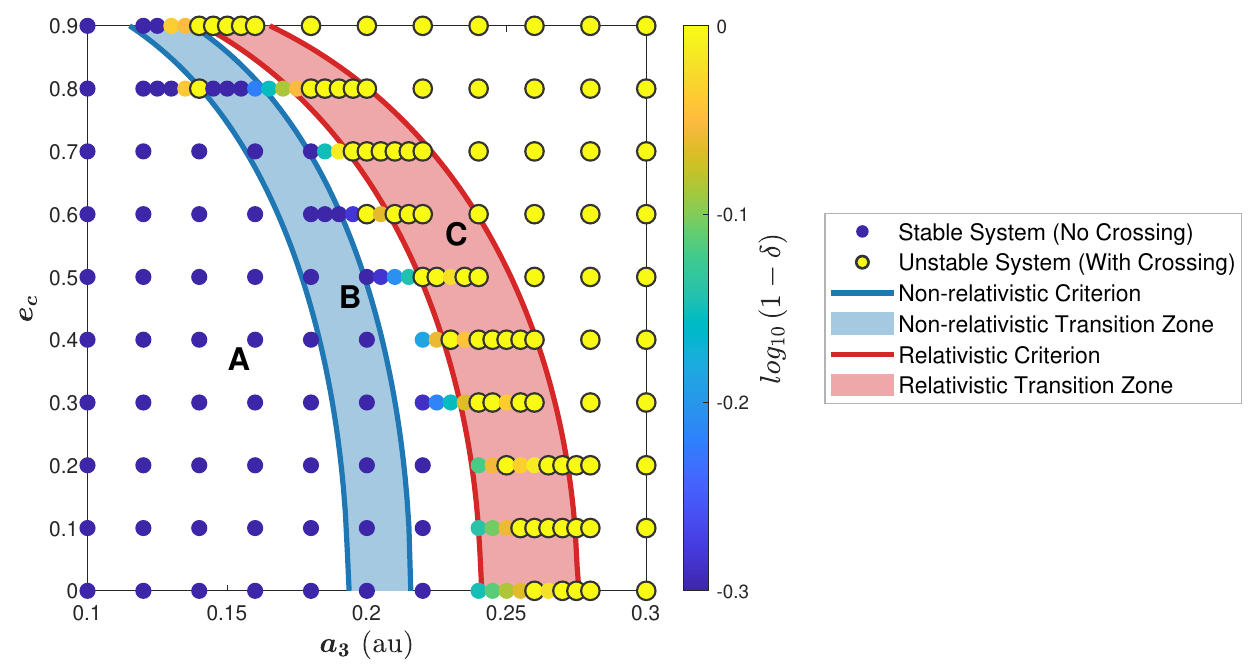}
    
    \caption{Stability criterion and simulation results for $3+1$ systems. As before, the stability criterion is compared with simulation results for $3+1$ systems. Relativistic and non-relativistic stability criterion based on Equation~(\ref{eq:e_crit 3+1}) are plotted in red and blue curves respectively. Each scattered points represents a numerically integrated system accounting for GR effects in the $e_c-a_3$ parameter space. Systems are integrated up to a maximum of $90$~Myr. The color code represents how close the orbit of the two planets become defined by $\log (1-\delta)$ (see Equation~(\ref{eq:delta})) throughout their evolution. Deep blue represents large separations between planet $2$ and planet $3$ without orbit crossing, whereas yellow stands for close encounters. The points with the black contours represent systems that experience orbit crossing. Here we plot all the values of $\log (1-\delta)$ at or below $-0.3$ in the same deepest blue color in order to show the transition from stable to unstable systems more clearly. In Figure \ref{fig:Evolution_3+1} we show the time evolution of three representative systems in different regions marked as A, B, and C in this plot.}
    
    \label{fig:Delta_3+1}
\end{figure*}

We simulate the evolution of systems with different initial conditions in the $e_c - a_3$ parameter space as a test of the derived criterion. Systems are integrated up to $90$~Myr or until orbit crossing. The simulation results are plotted in Figure~\ref{fig:Delta_3+1}. The same legend in Figure~\ref{fig:Delta_2+1} applies here. Each system is represented as a scattered point, and their colors reflect the closest distance between planet $3$ and its nearest neighbor planet $2$ throughout the evolution. A minimum cutoff value of $-0.3$ is imposed on $\log\left(1-\delta\right)$ in color representation to show the transition from stable to unstable more clearly. The relativistic and non-relativistic critical eccentricities are plotted in red and blue curves in Figure~\ref{fig:Delta_3+1} respectively. Systems lying below the red curves should be stable, while the ones above them should be unstable. 

\begin{figure*}
    \centering
    \includegraphics[width = \textwidth]{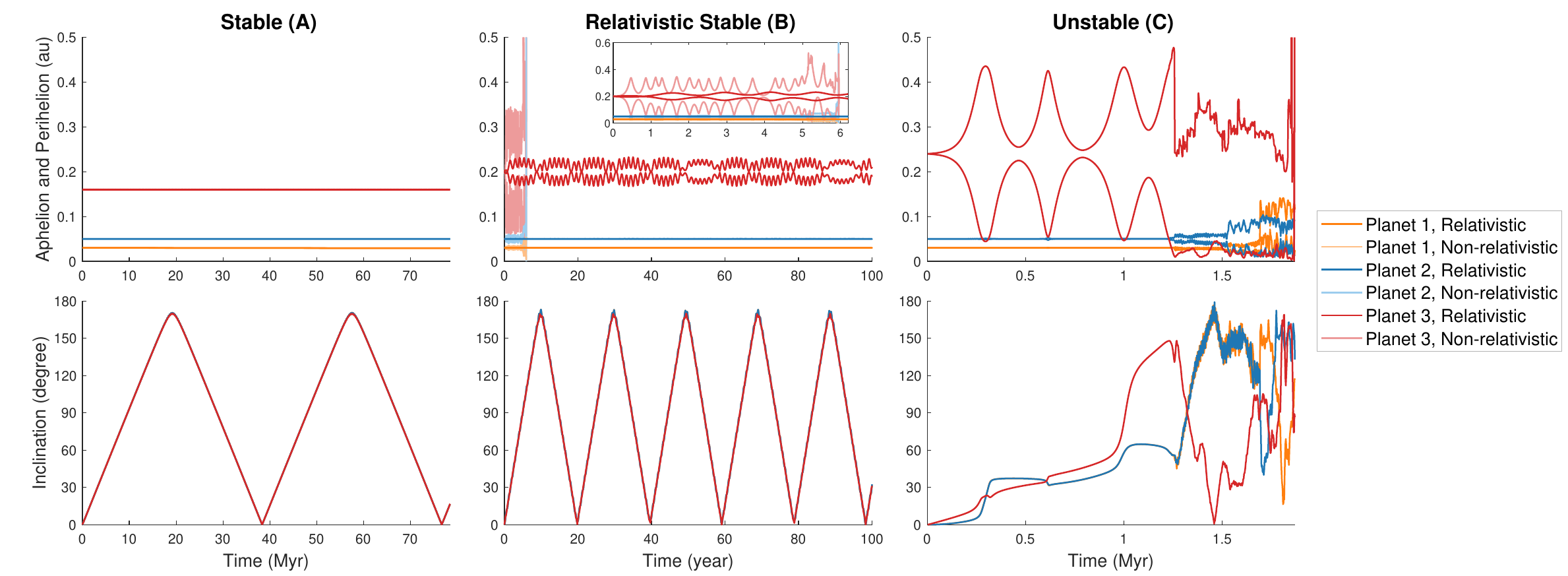}
    \caption{Time evolution of three representative systems. We show the time evolution of aphelion and perihelion (top row) and inclination (bottom row) for planet $1$ (solid orange), planet $2$ (solid blue), and planet $3$ (solid red) in three representative relativistic systems labeled as A, B, and C in Figure~\ref{fig:Delta_3+1}. For system B, non-relativistic evolution of aphelion and perihelion are overplotted in light blue and light red for comparison. Inset figure in the top-middle panel zoom in the first $2$~Myr to illustrate the system behavior of the non-relativistic case. The three systems share the following initial orbital parameters in common: $M=1M_\odot$, $m_1=m_2=m_3=10^{-6}M_\odot$, $m_c=10^{-3}M_\odot$, $a_1=0.03$ au, $a_2=0.05$ au, $a_c=5$ au, $e_1=e_2=e_3=0.001$, $i_1=i_2=i_3=0.001 rad$, $i_c=85^\circ$, and $\omega_1=\omega_2=\omega_3=\omega_c=\Omega_1=\Omega_2=\Omega_3=\Omega_c=0$. The only difference lies in $a_3$ and $e_c$. For the stable system A, $a_3=0.16$, $e_c=0.4$. For relativistic stable system B (only stable when considering GR), $a_3=0.2$, $e_c=0.5$. Finally, for unstable system C, $a_3=0.24$, $e_c=0.6$.}
    
    \label{fig:Evolution_3+1}
\end{figure*}

Most of the systems' behaviors agree with our expectations. It is worth mentioning that the simulation results conform with the criterion better than the $2+1$ case, mainly because the precession rates of planet $3$ are faster due to the presence of a closer neighboring planet $2$, as can be seen from Figure~\ref{fig:precession_2+1} and Figure~\ref{fig:precession_3+1}. The limitation on integration times are therefore mitigated. An exception at $a_3=0.14$ and $e_c=0.8$ is investigated in the the Appendix~\ref{appenix}. We expect that this system undergoes a resonant-like behavior because of equal precession rates from all three effects (see Figure~\ref{fig:precession_3+1}). GR precession tends to destabilize the resonant angle, inducing eccentricity excitations \citep[e.g.,][]{Naoz_2013a, Hansen_2020}. Notice that within the relativistic transition zone, only part of the systems experience ``true'' orbit crossing, while others all have very close orbital distances (as close as $5\times10^{-3}$~au), suggesting that longer integration timescale may indeed yield instability. This is consistent with the notion that systems in the transition zone will experience large eccentricity excitations and are on the brink of instability, but orbit crossing may happen very late in their evolution, exceeding our simulation capabilities.

Similarly to the $2+1$ case, in Figure~\ref{fig:Evolution_3+1} we plot the time evolution of three representative systems A, B, and C marked in Figure~\ref{fig:Delta_3+1}, corresponding to stable, relativistic stable, and unstable cases. Relativistic and non-relativistic criterion disagrees on the behavior of system B, and it is indeed the case according to the top middle panel and the zoomed-in over-plot in Figure~\ref{fig:Evolution_3+1}. Without GR, the eccentricity of planet $3$ oscillates largely that the orbits of inner planets cross and finally collide with the star (planet $1$) or become ejected out of the system (planet $2$) after only about $6$~Myr. In comparison, once GR is taken into account, the same system behaves stable over timescales of $100$~Myr with no close-encounters. Again, this is a convincing proof of our analytical criterion.

\begin{figure*}
    \centering
    \includegraphics[width = \textwidth]{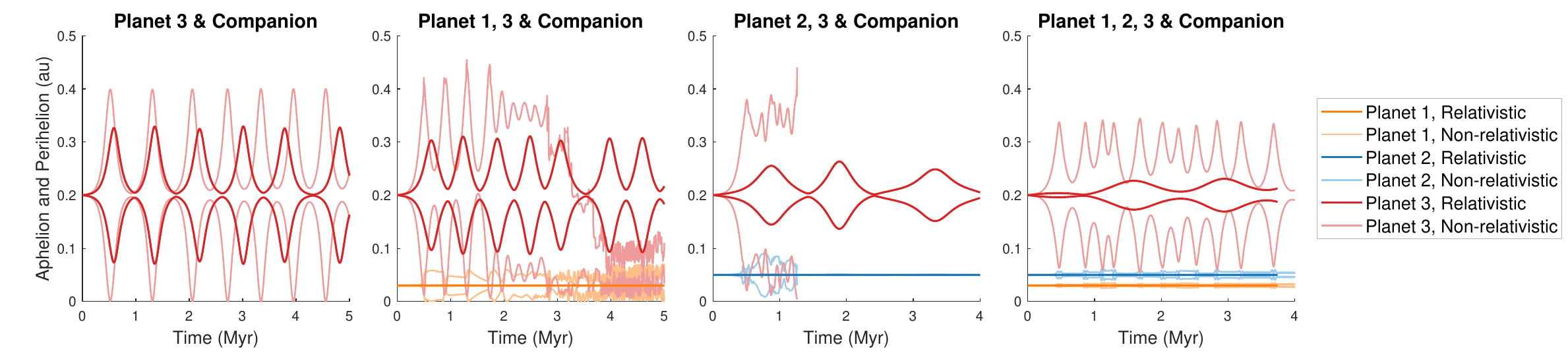}
    \caption{From left to right, we systematically add planet $1$ at $a_1=0.03$ au, planet $2$ at $a_2=0.05$ au, and both planet $1$ and $2$ into a system with initial conditions listed in Table~\ref{tab:3+1 initial conditions} with $a_3=0.2$ au and $e_c=0.5$. It is clear by comparison of different panels how the existence of planet $1$ and $2$ can help suppress the EKL eccentricity excitation on planet $3$ via Laplace-Lagrange secular interactions. Additionally, in each panel, the eccentricity excitation with GR (solid color lines) is much less than that of without GR (light color lines). In fact, GR helps prevent orbit crossing in an non-relativistic unstable system.}
    \label{fig:Evolution_i+1}
\end{figure*}

To illustrate the stabilizing effects of the Laplace-Lagrange and GR interactions comprehensively, we present another set of time evolution results of one of the $3+1$ systems initialized in Table~\ref{tab:3+1 initial conditions}, with the choice of $a_2=0.05$~au and $e_3=0.5$.
As shown in Figure~\ref{fig:Evolution_i+1}, we systematically add one or more planets into the system and inspect their influence on the eccentricity excitation of planet $3$. Specifically, from left to right, we begin with the configuration including the star, planet $3$, and the companion. Then in the second panel, we add planet $1$ only, followed by adding planet $2$ only in the third panel. Lastly, in the fourth panel, we consider the full $3+1$ system. The lines in solid colors represent the evolution in the presence of relativistic precession, while the lines in light colors represent the non-relativistic case. As shown Figure~\ref{fig:Evolution_i+1}, each time one or more planets are added into the system, the eccentricity oscillation is suppressed. The closer the added planet is, or for a higher multiple of planets, the larger the suppression of the eccentricities is. This trend is a direct reflection of the stabilizing effect of the Laplace-Lagrange interaction. Additionally, in each panel, the eccentricity excitation of the relativistic case (solid color lines) is always much smaller than the non-relativistic case (light color line), highlighting the efficiency of GR precession in suppressing the eccentricity excitations induced by the EKL effect from the companion. A combined effect of Laplace-Lagrange interactions and short-range forces or effects such as general relativity contributes to stabilizing observed systems against a faraway companion.

\section{Application on observed systems}
\label{sec:application}

The stability criterion developed in Section~\ref{sec:criterion} can be used to constrain the space of orbital parameters in which a hidden companion may reside \citep[][]{Faridani_2021}. As an application, we apply the criterion to two representative observed two-planet systems and identify the parameter space which an undiscovered companion may inhabit. Unlike Figure~\ref{fig:Delta_2+1}, here we focus on the companion's $e_c-a_c$ parameter space. The companion's critical eccentricity $e_{c, {\rm crit}}$ as a function of $a_c$ divides the parameter space into stable and unstable regions. The systems are then numerically integrated in the presence of an additional companion spanning across the $e_c-a_c$ parameter space to test our prediction. As before, the companion is assumed to have one Jupiter mass $M_\mathrm{J}$ with an inclination of $85^{\circ}$ with respect to the first planet in the system. K2-223 and K2-229 are the systems of interest, since they both comprise ultra-short period planets whose GR induced precession rates are faster than that of Laplace-Lagrange. Note that \citet{Faridani_2021} compared the GR and Laplace-Lagrange precession rates for all observed two-planet systems orbiting single stars with both planet masses known. They found that for a certain fraction of the systems, GR precession may contribute to their stabilization. Here, we are selecting two specific systems that do not require a too far away companion to distinguish the non-GR and GR criterion, thus less computationally expensive in terms of integration time.

K2-223 is a $1.06 M_\odot$ star orbited by two planets of $0.9 M_\oplus$ and $5.0 M_\oplus$ with semi-major axes of $0.0127$ au and $0.0549$ au respectively \citep[][]{Livingston_2018}. K2-229 is a K-type main sequence star of $0.837 M_\odot$, orbited by two confirmed planets of $2.59 M_\oplus$ and $21.3 M_\oplus$, at distances of $0.012888$ au and $0.07577$~au \citep[][]{Santerne_2018}\footnote{Unconfirmed candidate planet K2-229 d is not considered here.}. Unavailable parameters are assumed to be zero or near zero values to avoid peculiar behaviors of the integrator. Initial conditions for both systems are listed in Table~\ref{tab:K2 initial conditions}.

\begin{table}
	\centering
	\caption{Initial conditions of K2-223 and K2-229 with a hypothetical companion. The parameters of stars and inner planets are obtained from \citet{Livingston_2018} for K2-223 and \citet{Santerne_2018} for K2-229.}
	\label{tab:K2 initial conditions}
	\begin{tabular}{lcccc} 
		\hline
		Object & Mass & $a$ (au) & $e$ & $i$\\
		\hline
		K2-223 & $1.06~M_\odot$ & & & \\
		K2-223 b & $0.9~M_\oplus$ & $0.0127$ & $10^{-3}$ & $10^{-3}$ rad\\
		K2-223 c & $5.0~M_\oplus$ & $0.0549$ & $10^{-3}$ & $10^{-3}$ rad\\
		Companion & $1.0~M_J$ & $0.5 - 2.0$ & $0 - 0.9$ & $85^{\circ}$\\
		\hline
		K2-229 & $0.837~M_\odot$ & & & \\
		K2-229 b & $2.59~M_\oplus$ & $0.012888$ & $10^{-3}$ & $10^{-3}$ rad\\
		K2-229 c & $21.3~M_\oplus$ & $0.07577$ & $10^{-3}$ & $10^{-3}$ rad\\
		Companion & $1.0~M_J$ & $0.5 - 2.5$ & $0 - 0.9$ & $85^{\circ}$\\
		\hline
	\end{tabular}
\end{table}

\begin{figure*}
    \centering
    \includegraphics[width = \textwidth]{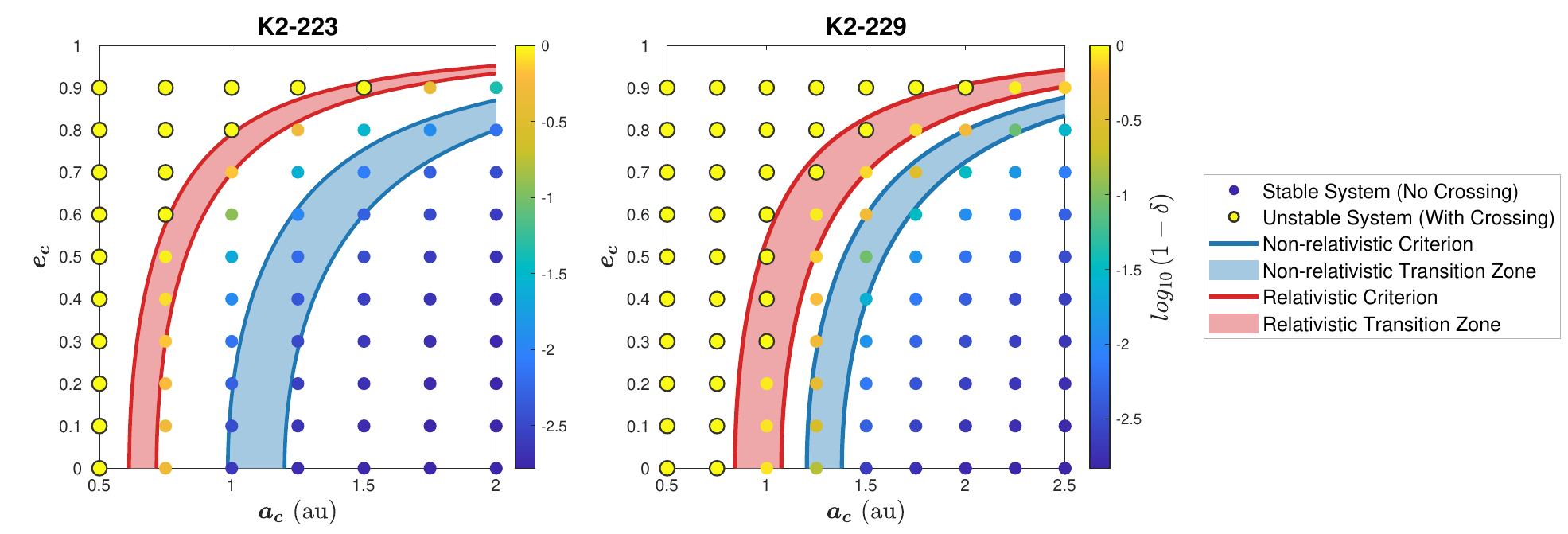}
    \caption{Application of stability criterion to observed systems. Stability criterion is applied to two observed systems, K2-223 and K2-229, to constrain the orbital elements of a possible undiscovered companion. We assume the presence of a hypothetical companion of $1 M_J$ with inclination $85^{\circ}$. Relativistic and non-relativistic criterion are plotted in red and blue respectively, dividing the $e_c-a_c$ space. The companion is prohibited to reside above the red curves, as the system would be unable to survive. As before, each scattered point represents a simulated system. The systems are integrated up to a maximum of $2.7$~Myr for K2-223 and $3.0$~Myr for K2-229. Stable systems are depicted in deep blue, while systems experiencing close-encounters are painted in yellow. Black contours around yellow points indicate orbit crossings. The boundary of black contours and the relativistic criterion agrees well.}
    \label{fig:application}
\end{figure*}

Figure~\ref{fig:application} illustrates the theoretical predictions of relativistic and non-relativistic criterion in the $e_c-a_c$ space in red and blue curves respectively. The enclosed light red and light blue areas depict the transition zones. Potential companions are prohibited to inhabit the parameter space above the red curves according to the relativistic stability criterion. Note that the different shape of the curves in Figure~\ref{fig:application}, compared to Figures \ref{fig:Delta_2+1} and \ref{fig:Delta_3+1}, is due to our focus on the companion's parameter space, rather than one orbital element of the inner planets. As before, each scattered point represents a corresponding numerically integrated system. The systems are integrated up to a maximum of $2.7$~Myr for K2-223 and $3.0$~Myr for K2-229. Stable systems with no orbit crossing are colored in deep blue, while unstable systems with orbit crossings are painted in bright yellow with a black contour. It is clear that the simulation results agree flawlessly with our analytical prediction. The successful application of the criterion developed in this work on observed systems enables a confident constraint on hidden inclined companions.

\section{Discussion}
\label{sec:discussion}

Multi-planet systems are a pervasive configuration in our Galaxy \citep[e.g.,][]{Tremaine_2012, Fabrycky_2014, Rowe_2014, Lissauer_2014}. Many of them are believed to be compact systems with tight planetary orbits \citep[e.g.,][]{Howard_2012, Brewer_2018, Weiss_2018, Winn_2018, Wu_2019, Weiss_2020}. The close distance to the star makes it possible for general relativity to shape their dynamical features. Meanwhile, stellar systems with faraway companions, either planets or stars, are also abundant \citep[e.g.,][]{Raghavan_2010, Knutson_2014, Konopacky_2016, Bryan_2016, Zhu_2018, Bryan_2019}. While these two populations are detected using different methods, existing evidence shows that multi-planet systems may be accompanied by faraway companions \citep[e.g.,][]{Lai_2017, Becker_2020}.

In this work, we have examined the orbital stability of hierarchical planetary systems with multiple nearly coplanar close-in planets and an inclined, distant, massive companion, i.e., the $N+1$ configuration (illustrated in Figure \ref{fig:Cartoon}). The companion tends to excite the eccentricities of inner planets and possibly drive the system to instability via the EKL mechanism \citep[see][]{Naoz_2016}. However, both the Laplace-Lagrange planet-planet interaction and the short-range interactions, such as relativistic precession induced by the star, can help suppress the EKL eccentricity excitations from the companion. 

We have developed a general analytical criterion that predicts the long-time stability of the system based on its initial conditions by comparing the precession rates from the relevant effects (see Equation~(\ref{eq:criterion})). Systems with fast EKL precession rates, compared to the aforementioned interactions, can be driven to instability. Conversely, the Laplace-Lagrange planet-planet interactions and short-range forces or effects, such as relativistic precession, can stabilize the system against eccentricity excitations if their precession rates are faster than that of EKL's. Similar to \citet{Denham_2018}, we have identified a transition zone in the parameter space between stable and unstable systems. Within the transition zone, the inner planets' eccentricities are excited to large values but they do not necessarily always undergo orbit crossing. 

Aside from the general relativity we consider here, other possible short-range interactions, such as tidal forces \citep[e.g.,][]{Hut_1981, Eggleton_1998, Eggleton_2001, Fabrycky_2007, Liu_2015} and magnetic interactions \citep[e.g.,][]{Ahuir_2021} between the planets and the star can also contribute to the stabilization of such systems. We believe that any extra short-range forces or effects in addition to GR, which we consider in this work, can be accounted for by incorporating corresponding precession rates in the $\omega_{\rm SR}$ term in the general criterion in Equation~(\ref{eq:criterion}).

Two different configurations are inspected in this work. First, we apply the criterion to a set of $2+1$ systems with different initial conditions where GR plays a non-negligible role, extending the criterion developed in \citet{Denham_2018} by including GR effects. This extension of the stability zone is depicted in Figure~\ref{fig:Delta_2+1}, where we plot the non-relativist and relativistic criteria in blue and red curves respectively. To validate our theory, a series of different systems in the $e_c-a_2$ parameter space are integrated numerically using a combination of N-body integration and Gauss's averaging method. Our simulation results conform well with our analytical predictions, though there is a slight difference due to our limited integration time. Time evolution of three example systems in Figure~\ref{fig:Evolution_2+1} also proves our theory. Specifically, the eccentricity excitation in the presence of GR is found to be significantly suppressed compared to the non-relativistic case (see case B in Figure \ref{fig:Evolution_2+1}). This effect contributes to stabilizing observed two-body exoplanet systems against perturbations from external companions \citep[][]{Faridani_2021}.

As a further generalization, the analytical criterion is then applied to a set of $3+1$ systems as another example. Simulations using the same numerical techniques show consistency between analytical criterion and numerically integrated results in the $e_c-a_3$ parameter space (see Figure~\ref{fig:Delta_3+1}). Time evolution of three example systems are presented in Figure~\ref{fig:Evolution_3+1} and authenticates the prediction of the analytical criterion again. Notice that only part of the systems in the transition zone undergo orbit crossing, while others experience close orbit encounters. This confirms the notion that systems in the transition zone are on the brink of instability.

Finally, the criterion is applied to two observed systems, K2-223 and K2-229, with a hypothetical companion of $1 M_J$ and an inclination of $85^{\circ}$ with respect to the first planet in the system. As shown in Figure~\ref{fig:application}, the simulation results are consistent with our analytical  stability criterion. The agreement suggests that the methodology outlined here (i.e., the analytical criterion) can be applied to a wide range of systems to constrain the configuration of possible companions, such as stars or planets.

\acknowledgments
We thank the anonymous referee for their comments. We also thank Dan Fabrycky for useful suggestions. 
L.W and S.N acknowledge the support from the Cross-Disciplinary Scholars in Science and Technology (CSST) program of University of California, Los Angeles. 
T.F and S.N. acknowledge partial support from the NSF through grant No. AST-1739160. S.N. thanks Howard and Astrid Preston for their generous support.

\appendix
\section{A Resonant-Like System}
\label{appenix}

\begin{figure*}[hbt!]
    \centering
    \includegraphics[width=\textwidth]{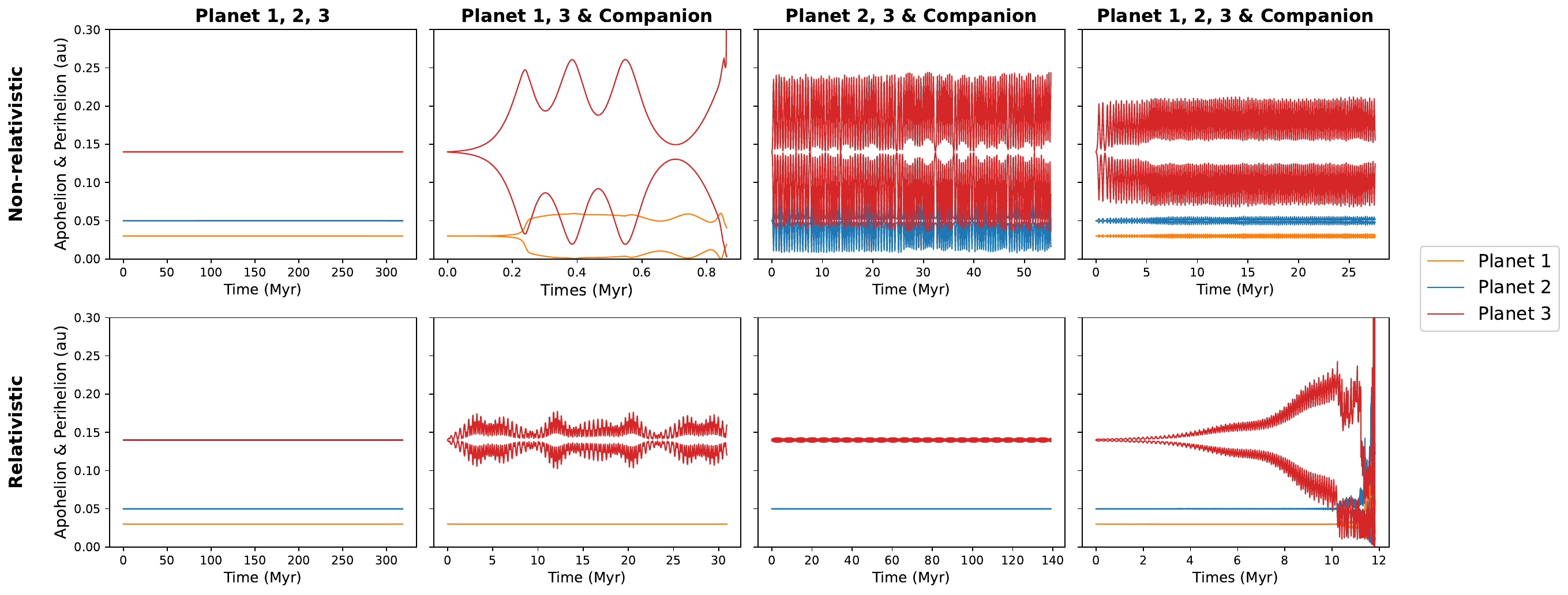}
    \caption{Illustration of a resonant system in Figure~\ref{fig:Delta_3+1} at $a_3=0.14$, $e_c=0.8$. We remove the companion, planet $2$, and planet $1$ in the first three columns respectively to identify the reason for the resonant behavior of the $3+1$ system. The last column shows the full $3+1$ configuration. The resonance behavior can only be observed when all planets and the companion are present, as well as GR effects.}
    \label{fig:appendix}
\end{figure*}
Figure~\ref{fig:Delta_3+1} shows a system that lies deep within the stable region but undergoes orbit crossing at $a_2=0.14$~au and $e_c=0.8$. As one can tell from Figure~\ref{fig:Delta_3+1}, its neighboring systems with either smaller or larger semi-major axis of planet $2$ are mostly stable. Together with its time evolution which is shown in Figure~\ref{fig:appendix}, we are convinced that this anomaly stems from a resonant-like behavior. In this Appendix, we show that this orbital crossing is a result of GR destabilizing the system. It was shown that GR precession can destabilize the resonant family, especially when the GR precession rate is similar to that of EKL, and sometimes even when the GR precession is shorter \citep[e.g.,][]{Ford_2000, Naoz_2013a, Hansen_2020}. According to Figure \ref{fig:precession_3+1}, the precession rate of EKL, GR, and Laplace-Lagrange are all comparable for planet $3$ at $a_2=0.14$~au and $e_c=0.8$, thus yielding a resonant-like behavior, as is suggested in the three-body case \citep{Naoz_2013a}. 

In Figure~\ref{fig:appendix}, we consider this configuration in both non-relativistic and relativistic perspectives (top and bottom rows, respectively) and systematically remove one different object from the system each time. Specifically, from left to right in Figure~\ref{fig:appendix}, we remove the companion, planet $2$, planet $1$, and present the full $3+1$ system respectively. One may wonder if the orbit crossing is a result of a Laplace-Lagrange resonance. In fact, as depicted in the first column of Figure~\ref{fig:appendix}, this system is stable in the absence of a faraway companion, both with and without GR precession, excluding the possibility of Laplace-Lagrange resonance among the inner planets. Analytical calculations of maximum eccentricities due to Laplace-Lagrange interactions alone also precludes this explanation. We then add either planet $1$ (second column) or planet $2$ (third column) to show that the GR precession, in fact, stabilizes the system in the $2+1$ case. Finally, under the full $3+1$ configuration, the resonant behavior only appears when GR is added, confirming the explanation that it is GR that brings about the resonance.


\bibliography{Reference}
\bibliographystyle{aasjournal}



\end{CJK*}
\end{document}